\documentclass[english]{smfart}

\def\CRAS{C.~R.~Acad.~Sc.~Paris}
\def\PTP{Prog.~Theor.~Phys.~}

\def \D {\hbox{d}}
\def\PI     {{\rm P1}}
\def\PII    {{\rm P2}}
\def\cKdV{c-KdV}

\def\KFIVone {K_{1,{\rm IV}}}
\def\KFIVtwo {K_{2,{\rm IV}}}

\def\QA{\tilde Q_1}
\def\QB{\tilde Q_2}
\def\PA{\tilde P_1}
\def\PB{\tilde P_2}
\def\Qj{\tilde Q_j}
\def\Pj{\tilde P_j}

\newcommand{\pard}[2]{\frac{\partial #1}{\partial #2}}     

\author[Conte]
{Robert CONTE}
\address{Service de physique de l'\'etat condens\'e (URA 2464), CEA--Saclay
\\ F--91191 Gif-sur-Yvette Cedex, France
}
\email{Robert.Conte@cea.fr}
\urladdr{}

\author[Musette]
{Micheline MUSETTE}
\address{Dienst Theoretische Natuurkunde, Vrije Universiteit Brussel
and
\\~~International Solvay Institutes for Physics and Chemistry
\\~~Pleinlaan 2, B--1050 Brussels, Belgium
}
\email{MMusette@vub.ac.be}
\urladdr{}

\author[Verhoeven]
{Caroline VERHOEVEN}
\address{Dienst Theoretische Natuurkunde, Vrije Universiteit Brussel
and
\\~~International Solvay Institutes for Physics and Chemistry
\\~~Pleinlaan 2, B--1050 Brussels, Belgium
}
\email{CVerhoev@vub.ac.be}
\urladdr{}

\title[H\'enon-Heiles Hamiltonians]
{Painlev\'e property of the H\'enon-Heiles Hamiltonians}
\alttitle{Propri\'et\'e de Painlev\'e des hamiltoniens de H\'enon-Heiles}

\thanks{
The authors acknowledge the financial support of
the Tournesol grant no.~T2003.09
between Belgium and France.
C.~Verhoeven is a postdoctoral fellow at the FWO-Vlaanderen.
}

\begin{document}

\frontmatter

\begin{abstract}
Time independent Hamiltonians of the physical type
$$H = (P_1^2+P_2^2)/2+V(Q_1,Q_2)$$
pass the Painlev\'e test for only seven potentials $V$,
known as the H\'enon-Heiles Hamiltonians,
each depending on a finite number of free constants.
Proving the Painlev\'e property was not yet achieved for generic values of
the free constants.
We integrate each missing case
by building a birational transformation to some fourth order first degree
ordinary differential equation in the classification (Cosgrove, 2000)
of such polynomial equations which possess the Painlev\'e property.
The properties common to each Hamiltonian are:\\
(i) the general solution is meromorphic and expressed with hyperelliptic functions
of genus two,\\
(ii) the Hamiltonian is complete (the addition of any time independent term
would ruin the Painlev\'e property).
\end{abstract}

\begin{altabstract}
Les hamiltoniens ind\'ependants du temps de la forme 
$$H = (P_1^2+P_2^2)/2+V(Q_1,Q_2)$$
satisfont au test de Painlev\'e pour seulement sept potentiels $V$~;
ceux-ci sont connus sous le nom de hamiltoniens de H\'enon-Heiles et ils 
d\'ependent d'un nombre fini de constantes libres.
La propri\'et\'e de Painlev\'e restait \`a \'etablir
 pour des valeurs g\'en\'eriques des constantes libres.
Nous traitons chacun des cas en suspens
en construisant une transformation birationnelle vers une 
\'equation diff\'erentielle ordinaire d'ordre quatre qui 
figure dans la liste exhaustive (Cosgrove, 2000)
de telles \'equations polynomiales
poss\'edant la propri\'et\'e de Painlev\'e.
Les propri\'et\'es communes \`a ces hamiltoniens sont~:\\
(i) la solution g\'en\'erale est m\'eromorphe et peut \^{e}tre exprim\'ee en termes de 
fonctions hyperelliptiques de genre deux,\\
(ii) le hamiltonien est complet au sens o\`u l'addition de tout terme
ind\'ependant du temps ferait perdre la propri\'et\'e de Painlev\'e.
\end{altabstract}


\altkeywords{
Hamiltonien d'H\'enon-Heiles,
propri\'et\'e de Painlev\'e,
hyperelliptique,
s\'eparation de variables,
coordonn\'ees de Darboux.
} 


\maketitle

\mainmatter


\section{Introduction}
\indent

Let us consider the most general
two-degree of freedom, classical, time-independent Hamiltonian
of the physical type (i.e, the sum of a kinetic energy and a potential energy),
\begin{eqnarray}
H
& =&
 \frac{1}{2} (p_1^2 + p_2^2) + V(q_1,q_2),
\label{eqH2TV}
\end{eqnarray}
and let us require that the general solution $q_1^{n_1}, q_2^{n_2}$,
with $n_1,n_2$ integers to be determined,
be single valued functions of the complex time $t$,
i.e., what is called the \textit{Painlev\'e property} of these equations.

A necessary condition is that the Hamilton equations of motion,
when written in these variables $q_1^{n_1}, q_2^{n_2}$,
pass the Painlev\'e test (\cite{Cargese1996Conte}).
This selects seven potentials $V$ (three ``cubic'' and four ``quartic'')
depending on a finite number of arbitrary constants,
which are known as the H\'enon-Heiles Hamiltonians (\cite{HH}).
In order to prove the sufficiency of these conditions,
one must then perform the explicit integration and check the
singlevaluedness of the general solution.
We present here a review on this subject.
\bigskip

The paper is organized as follows:
\bigskip

In section \ref{sectionThe_seven},
we enumerate the seven cases isolated by the Painlev\'e test,
together with the second constant of the motion $K$
in involution with the Hamiltonian.
In section \ref{sectionCasesWithSeparatingVariables},
we recall the separating variables in the four cases where they are known.
In section \ref{sectionConfluences},
we display confluences from quartic cases to all the cubic cases,
thus restricting the problem to the consideration of the quartic cases only.
In section \ref{sectionFourthOrderODEs},
due to the lack of knowledge of the separating variables
in the three remaining cases,
we state the equivalence of the equations of motion and the conservation
of energy
with some fourth order first degree ordinary differential equations (ODEs).
In section \ref{sectionBirational},
since these fourth order equations do not belong to any set of already classified
equations,
we build a birational transformation between each quartic case
and some fourth order ODE belonging to
a classification of Cosgrove (\cite{Cos2000a}),
thus proving the Painlev\'e property for the quartic cases.
\bigskip

To summarize, the results are twofold:
\begin{enumerate}
\item
each case is integrated by solving a Jacobi inversion problem
involving a hyperelliptic curve of genus two,
which proves the meromorphy of the general solution,
\item
each case is \textit{complete} in the sense of Painlev\'e,
i.e, it is impossible to add any time-independent term to the Hamiltonian
without ruining the Painlev\'e property.
\end{enumerate}

\section     {The seven H\'enon-Heiles Hamiltonians}
\label{sectionThe_seven}
\indent

By application of the Painlev\'e test,
one isolates two classes of potentials $V(q_1,q_2)$,
called ``cubic'' and ``quartic'' for simplification.
\bigskip

\begin{enumerate}
\item
In the cubic case HH3
(\cite{CTW,Fordy1991,CFP1993}),
       \index{H\'enon-Heiles Hamiltonian!cubic}
\begin{eqnarray}
H
& =&
 \frac{1}{2} (p_1^2 + p_2^2 + \omega_1 q_1^2 + \omega_2 q_2^2)
    + \alpha q_1 q_2^2 - \frac{1}{3} \beta q_1^3
    + \frac{1}{2} \gamma q_2^{-2},\quad
\alpha \not=0,
\label{eqHH0}
\end{eqnarray}
\noindent in which the constants $\alpha,\beta,\omega_1,\omega_2$ and $\gamma$
can only take the three sets of values,
\begin{eqnarray}
\hbox{(SK)} : & & \beta/ \alpha=-1, \omega_1=\omega_2,\
\label{eqHH3SKcond} \\
\noalign{\medskip}
\hbox{(KdV5)} : & & \beta/ \alpha=-6,\
\label{eqHH3K5cond}\\
\noalign{\medskip}
\hbox{(KK)} : & & \beta/ \alpha=-16, \omega_1=16 \omega_2.
\label{eqHH3KKcond}
\end{eqnarray}
\medskip
\item
In the quartic case HH4
(\cite{RDG1982,GDR1983}),
\index{H\'enon-Heiles Hamiltonian!quartic}
\begin{eqnarray}
\begin{array}{ccl}H & = &\displaystyle 
\frac{1}{2}(P_1^2+P_2^2+\Omega_1 Q_1^2+\Omega_2 Q_2^2)
 +C Q_1^4
\\
\noalign{\medskip}
& &\displaystyle 
\qquad\qquad\qquad  + B Q_1^2 Q_2^2 + A Q_2^4+\frac{1}{2}\left(\frac{\alpha}{Q_1^2}+\frac{\beta}{Q_2^2}\right)
 + \gamma Q_1,\quad B \not=0,
 \end{array}
\label{eqHH40}
\end{eqnarray}
\noindent in which the constants
$A,B,C,\alpha,\beta,\gamma,\Omega_1$ and $\Omega_2$
can only take the four values
(the notation $A:B:C=p:q:r$ stands for $A/p=B/q=C/r=\hbox{arbitrary}$),
\begin{eqnarray}
& & \left\lbrace
\begin{array}{ll}
\displaystyle{
A:B:C=1:2:1,\ \gamma=0,
}
\\
\noalign{\medskip}
\displaystyle{
A:B:C=1:6:1,\ \gamma=0,\ \Omega_1=\Omega_2,
}
\\
\noalign{\medskip}
\displaystyle{
A:B:C=1:6:8,\ \alpha=0,\ \Omega_1=4\Omega_2,
}
\\
\noalign{\medskip}
\displaystyle{
A:B:C=1:12:16,\ \gamma=0,\ \Omega_1=4\Omega_2.
}
\end{array}
\right.
\label{eqHH4NLScond}
\end{eqnarray}
\end{enumerate}
\bigskip

For each of the seven cases so isolated
there exists a second constant of the motion $K$
(\cite{Drach1919KdV,BEF1995b,H1984}) 
(\cite{H1987,BakerThesis,BEF1995b}) 
in involution with the Hamiltonian,

$$ \, K\ \ =\ 
\left(3 p_1 p_2 + \alpha q_2 (3 q_1^2 + q_2^2) + 3 \omega_2 q_1 q_2 \right)^2
+ 3 \gamma (3 p_1^2 q_2^{-2} + 4 \alpha q_1 + 2 \omega_2),\ \leqno{(SK)}
\label{eqHH3SKSecond}
$$
$$\begin{array}{ccl}
 K & =& 
4 \alpha p_2 (q_2 p_1 - q_1 p_2)
+ (4 \omega_2 - \omega_1) (p_2^2 + \omega_2 q_2^2 + \gamma q_2^{-2})\qquad\qquad\ 
\\ 
\noalign{\medskip}
& &
\hfill+ \alpha^2 q_2^2 (4 q_1^2 + q_2^2)
+ 4 \alpha q_1 (\omega_2 q_2^2 - \gamma q_2^{-2}),
\end{array}\leqno{(KdV5)}
\label{eqHH3K5Second}
$$
$$\begin{array}{ccl}
 K & =& 
(3 p_2^2 + 3 \omega_2 q_2^2 + 3 \gamma q_2^{-2})^2
+ 12 \alpha p_2 q_2^2 (3 q_1 p_2 - q_2 p_1)\qquad\qquad\ 
 \\
\noalign{\medskip}
  & &
\hfill- 2 \alpha^2 q_2^4 (6 q_1^2 + q_2^2)
+ 12 \alpha q_1 (-\omega_2 q_2^4 + \gamma)
- 12 \omega_2 \gamma,
\end{array}\leqno{(KK)}
\label{eqHH3KKSecond}$$
$$\left\lbrace\begin{array}{ccl} K &=&\displaystyle  
\left(Q_2 P_1 - Q_1 P_2 \right)^2
+ Q_2^2 \frac{\alpha}{Q_1^2} + Q_1^2 \frac{\beta}{Q_2^2}\\
&&\hspace{-.8cm}\displaystyle -\frac{\Omega_1-\Omega_2}{2}
\left(P_1^2-P_2^2+Q_1^4-Q_2^4+\Omega_1 Q_1^2 - \Omega_2 Q_2^2
+ \frac{\alpha}{Q_1^2} - \frac{\beta}{Q_2^2}
 \right),\\
\noalign{\medskip}
 A&=&\displaystyle \frac{1}{2},
 \end{array}\right.\leqno{(1:2:1)}
\label{eqHH401:2:1K}
$$
\smallskip

$$\left\lbrace\begin{array}{ccl} K &=&\displaystyle 
\left(
P_1 P_2 + Q_1 Q_2 \Big(-\frac{Q_1^2+Q_2^2}{8}+\Omega_1 \Big)
\right)^2\\
\noalign{\medskip}
  & &\displaystyle 
- P_2^2 \frac{\kappa_1^2}{Q_1^2}
- P_1^2 \frac{\kappa_2^2}{Q_2^2}
+\frac{1}{4}\left(\kappa_1^2 Q_2^2 + \kappa_2^2 Q_1^2 \right)
+\frac{\kappa_1^2 \kappa_2^2}{Q_1^2 Q_2^2},\
\\
\noalign{\medskip}
&&\displaystyle \hspace{-1cm}\alpha=-\kappa_1^2,\ \ 
\beta= - \kappa_2^2,\ \ A=-\frac{1}{32},
 \end{array}\right.\leqno{(1:6:1)}
$$
\smallskip

$$\left\lbrace\begin{array}{ccl} K &=&\displaystyle 
\left(
P_2^2-\frac{Q_2^2}{16}(2 Q_2^2+4 Q_1^2+\Omega_2)
     +\frac{\beta}{Q_2^2}
\right)^2
\\
\noalign{\medskip}
& &\displaystyle 
-\frac{1}{4}Q_2^2(Q_2 P_1 - 2 Q_1 P_2)^2+\gamma
\bigg(
-2 \gamma Q_2^2
-4 Q_2 P_1 P_2
\\
\noalign{\medskip}
& &\displaystyle 
\hfill 
+\frac{1}{2} Q_1 Q_2^4+ Q_1^3 Q_2^2
+4 Q_1 P_2^2 -4 \Omega_2 Q_1 Q_2^2
+ 4 Q_1 \frac{\beta}{Q_2^2}
\bigg),\\
\noalign{\medskip}
A&=&\displaystyle -\frac{1}{16},
 \end{array}\right.\leqno{(1:6:8)}
$$
\smallskip

$$\left\lbrace\begin{array}{ccl} K &=&\displaystyle 
\bigg(8 (Q_2 P_1 - Q_1 P_2) P_2 - Q_1 Q_2^4 - 2 Q_1^3 Q_2^2\\
\noalign{\medskip}
&& \displaystyle + 2 \Omega_1 Q_1 Q_2^2 - 8 Q_1 \frac{\beta}{Q_2^2} \bigg)^2
+\frac{32 \alpha}{5}
\left(Q_2^4 + 10 \frac{Q_2^2 P_2^2}{Q_1^2}\right),\\
\noalign{\medskip}
A&=&\displaystyle -\frac{1}{32}.
 \end{array}\right.\leqno{(1:12:16)}
$$
\bigskip

\textit{Remark}.

Performing the reduction $q_1=0,p_1=0$ in the three HH3 Hamiltonians (\ref{eqHH0})
yields $H=p^2/2 + (1/2) \omega q^2 + (1/2) \gamma q^{-2}$,
for which $q^2$ obeys a linearizable Briot-Bouquet ODE.
Similarly, the reduction $Q_1=1,P_1=0$ in the four HH4 Hamiltonians (\ref{eqHH40})
yields $H=P^2/2+(1/2) \omega Q^2 + A Q^4 + (1/2) \beta Q^{-2}$,
for which $Q^2$ obeys the Weierstrass elliptic equation.
\medskip

These seven H\'enon-Heiles Hamiltonians
can be studied from various points of view such as:
separation of variables (\cite{Sklyanin1995}),
Painlev\'e property,
algebraic complete integrability (\cite{AvM1989}).
For the interrelations between these various approaches,
the reader can refer to the plain introduction in Ref.~\cite{AMB}.
In the present work, we only deal with proving the Painlev\'e property (PP).
\smallskip

In order to prove or disprove the PP,
it is sufficient to obtain an (explicit) canonical transformation
to new canonical variables
(the so-called \textit{separating variables})
which separate the
\textit{Hamilton-Jacobi equation} for the action $S(q_1,q_2)$
(\cite[chap.~10]{ArnoldMechanics}),
 \index{Hamilton-Jacobi equation}
which for two degrees of freedom is
\begin{eqnarray}
& &
H(q_1,q_2,p_1,p_2)-E=0,\
p_1=\pard{S}{q_1},\
p_2=\pard{S}{q_2}.
\label{eqHJAutonomous}
\end{eqnarray}
Indeed, if such separating variables are obtained,
depending on the genus $g$ of the hyperellitic curve $r^2=P(s)$
involved in the associated Jacobi inversion problem,
\begin{eqnarray}
& &
\frac{    \D s_1}{\sqrt{P(s_1)}}+\frac{     \D s_2}{\sqrt{P(s_2)}}=0,
\qquad
\frac{s_1 \D s_1}{\sqrt{P(s_2)}}+ \frac{s_2 \D s_2}{\sqrt{P(s_2)}}= \D t,
\label{eqJacobiInversionProblem}
\end{eqnarray}
the elementary symmetric functions $s_1+s_2$ and $s_1 s_2$
are either meromorphic functions of time ($g \le 2$),
or multivalued ($g >3$).

\section     {The four cases with known separating variables}
\label{sectionCasesWithSeparatingVariables}
\indent

Two of the seven cases (KdV5, 1:2:1) have a
second invariant $K$ equal to a second degree polynomial in the momenta,
therefore there exists a classical method (\cite{Staeckel1893a,Staeckel1893b})
to obtain the canonical transformation
$(q_1,q_2,p_1,p_2) \to (s_1,s_2,r_1,r_2)$
with the separating variables
$(s_1,s_2)$ obeying the canonical system (\ref{eqJacobiInversionProblem}).
For the KdV5 case, one obtains
(\cite{Drach1919KdV,AP1983,Woj1984})
\begin{eqnarray}
& &
\left\lbrace
\begin{array}{ll}
\displaystyle{
q_1 = -(s_1+s_2+\omega_1-4 \omega_2)/(4 \alpha),\quad
q_2^2=- s_1 s_2 /(4 \alpha^2),\
}
\\
\noalign{\medskip}
\displaystyle 
p_1=-4 \alpha\frac{s_1 r_1 - s_2 r_2}{s_1-s_2},\qquad\qquad\qquad
p_2^2=-16 \alpha^2 \frac{s_1 s_2 (r_1-r_2)^2}{(s_1-s_2)^2},
\\
\noalign{\medskip}
\displaystyle{
H=\frac{f(s_1,r_1)-f(s_2,r_2)}{s_1-s_2},\
}
\\
\noalign{\medskip}
\displaystyle{
f(s,r)=-\frac
  {s^2 (s+\omega_1-4 \omega_2)^2 (s-4 \omega_2)-64 \alpha^4 \gamma}{32 \alpha^2 s}
  +8 \alpha^2 r^2 s,
}
\\
\noalign{\medskip}
\displaystyle{
f(s_j,r_j)- E s_j + \frac{K}{2}=0,\ j=1,2,
}
\\
\noalign{\medskip}
\displaystyle{
P(s)=s^2 (s+\omega_1-4 \omega_2)^2 (s-4 \omega_2)
       + 32 \alpha^2 E s^2 - 16 \alpha^2 K s -64 \alpha^4 \gamma.
}
\end{array}
\right.
\end{eqnarray}
For 1:2:1, one obtains 
\begin{eqnarray}
& &
\left\lbrace
\begin{array}{rcl}
q_j^2&=&\displaystyle (-1)^j \frac{(s_1+\omega_j)(s_2+\omega_j)}{\omega_1-\omega_2},\ \ j=1,2,
\\
\noalign{\bigskip}
p_j&=&\displaystyle 2 q_j \frac{\omega_{3-j} (r_2-r_1) -s_1 r_1 + s_2 r_2}
{s_1-s_2},\ \ j=1,2,
\\
\noalign{\bigskip}
H&=&\displaystyle \frac{f(s_1,r_1)-f(s_2,r_2)}{s_1-s_2},\
\\
\noalign{\bigskip}
f(s,r)&=& \displaystyle 
2(s+\omega_1)(s+\omega_2) r^2 - \frac{s^3}{2}-\frac{\omega_1+\omega_2}{2}s^2\\
\noalign{\medskip}
&&\displaystyle\hfill-\frac{\omega_1 \omega_2}{2} s
+\frac{\omega_2- \omega_1}{2}
\Big(\frac{\alpha}{s+\omega_1}-\frac{\beta}{s+\omega_2}\Big),
\\
\noalign{\bigskip}
f(s_j,r_j) &=&\displaystyle - \Big(s_j + E \frac{\omega_1+\omega_2}{2} \Big)
- \frac{\alpha+\beta}{2}- \frac{K}{2},\ \ j=1,2,
\\
\noalign{\bigskip}
\displaystyle
P(s)&=&\displaystyle
s(s+\omega_1)^2(s+\omega_2)^2
-\alpha (s+\omega_2)^2-\beta (s+\omega_1)^2
\\
\noalign{\smallskip}
&&\displaystyle\hfill-(s+\omega_1)(s+\omega_2)
\left[E (2s+\omega_1+\omega_2)-K\right].
\end{array}
\right.
\end{eqnarray}

The two cubic cases SK and KK,
\begin{eqnarray}
H_{\rm SK}
& =&
 \frac{1}{2} (P_1^2 + P_2^2)  + \frac{\Omega_1}{2} (Q_1^2 + Q_2^2)
+ \frac{1}{2} Q_1 Q_2^2 + \frac{1}{6} Q_1^3 + \frac{\lambda^2}{8} Q_2^{-2},\
\label{eqHH3SK}
\\
\noalign{\smallskip}
H_{\rm KK}
& =&
 \frac{1}{2} (p_1^2 + p_2^2) + \frac{\omega_2}{2} (16 q_1^2 + q_2^2)
    + \frac{1}{4} q_1 q_2^2 + \frac{4}{3} q_1^3 + \frac{\lambda^2}{2} q_2^{-2},
\label{eqHH3KK}
\end{eqnarray}
are
equivalent under a birational canonical transformation
(\cite{BW1994,SEL}).
Therefore, the separating variables $(s_1,s_2)$ are common to these two cases.

In the nongeneric case $\lambda=0$, the separating variables have been built (\cite{RGC})
by a method (\cite{AvM1987,VanhaeckeLNM})
based on the local representation of the general solution $q_1(t),q_2(t)$
by a Laurent series of $t-t_0$ near a movable singularity $t_0$.
The algebraic curves defined by the values of the two invariants $H,K$
in terms of the arbitrary coefficients of the Laurent series
are then geometrically interpreted,
with, in principle, the separating variables as the final output.
However,
some technical difficulty prevents this method to handle the generic case
$\lambda\not=0$.

The generic case can nevertheless be separated (\cite{VMC2002a}) and the result is

\begin{eqnarray}
& &
\left\lbrace
\begin{array}{ll}
\displaystyle{
q_1=-6 \left(\frac{\PA - \PB}{\QA - \QB} \right)^2 -\frac{\QA + \QB}{2},
\qquad\quad
q_2^2=24 \frac{f(\QA,\PA)-f(\QB,\PB)}{\QA - \QB},
}
\\
\noalign{\bigskip}
\displaystyle{
p_1=-4 \QA \frac{\PA - \PB}{\QA - \QB} - 2 \frac{\QA \PB - \QB \PA}{\QA - \QB},\
\quad p_2= \QB \frac{\PA - \PB}{\QA - \QB},
}
\\
\noalign{\bigskip}
\displaystyle{
H=f(\QA,\PA)+f(\QB,\PB)
 + \frac{\lambda^2}{24} \frac{\QA-\QB}{f(\QA,\PA)-f(\QB,\PB)},
}
\\
\noalign{\bigskip}
\displaystyle{
f(q,p)=p^2+\frac{1}{12}q^3-4 \omega_2^2 q,
}
\\
\noalign{\bigskip}
\displaystyle{
\left(f(\Qj,\Pj) - \frac{E}{2}\right)^2 + \frac{\lambda^2}{24} \Qj + K=0,\quad j=1,2,
}
\\
\noalign{\bigskip}
\displaystyle{
\QA=s_1^2- \frac{3 K}{\lambda^2},\
\QB=s_2^2- \frac{3 K}{\lambda^2},\
\PA=\frac{r_1}{2 s_1},\
\PB=\frac{r_2}{2 s_2},\
}
\\
\noalign{\bigskip}
\displaystyle{
P(s)=-\frac{1}{3}\left(s^2-3 \frac{K}{\lambda^2}\right)^3
    + \Omega_1^2 \left(s^2-3 \frac{K}{\lambda^2}\right)
 + \frac{\lambda}{\sqrt{3}} s + 2 E.
}
\end{array}
\right.
\end{eqnarray}
\bigskip

It is remarkable that the canonical transformation
\begin{eqnarray}
& &
(q_1,q_2,p_1,p_2) \to
\left(\frac{\QA+\QB}{2}+ \Omega_1,\frac{\QA-\QB}{2},\PA+\PB,\PA-\PB\right)
\end{eqnarray}
coincides with the canonical transformation between
the SK variables and the KK variables
in the particular case $\lambda=0$.
\bigskip

In the three remaining cases, the quartic 1:6:1, 1:6:8, 1:12:16,
the separating variables are only known in nongeneric cases
(\cite{V2003,VMC2003}),
and the associated \textit{particular} solutions are single valued.
In order to decide about the Painlev\'e property,
which only involves the \textit{general} solution,
one must therefore integrate by different means.

\section     {Confluences from the quartic cases to the cubic ones}
\label{sectionConfluences}
\indent

A possible way to integrate would be to take advantage of some confluence
from an integrated case to a not yet integrated case.
For instance,
the property of single valuedness of the general solution
of the second Painlev\'e equation $\PII$ implies,
from the classical confluence from $\PII$ to $\PI$,
the same property for $\PI$.

The confluence from the quartic 1:6:8 case to the cubic KK case
found in Ref.~(\cite{Rom1995b})
is not an isolated feature (\cite{V2003}),
and in fact all the cubic cases can be obtained
by a confluence of at least one quartic case.
Just like between the six Painlev\'e equations,
one of the parameters in the Hamiltonian is lost in the process.
Consider, for instance, the quartic 1:12:16 and the cubic KK cases,
\begin{eqnarray}
& &
\left\lbrace
\begin{array}{ll}
\displaystyle{
h_{1:12:16}(t) =
\frac{1}{2}(p_1^2+p_2^2) + \frac{\omega}{8} (4 q_1^2+ q_2^2)\qquad\qquad\qquad
}
\\
\noalign{\medskip}
\displaystyle{
\phantom{1234}
\hfill\qquad\qquad\qquad\qquad - \frac{n}{32} (16 q_1^4+ 12 q_1^2 q_2^2 + q_2^4)
 +\frac{1}{2}\left(\frac{\alpha}{q_1^2}+\frac{\beta}{q_2^2}\right),
}
\\
\noalign{\bigskip}
\displaystyle{
H_{\rm KK}(T)=
 \frac{1}{2} (P_1^2 + P_2^2) + \frac{\Omega}{2} (16 Q_1^2 + Q_2^2)
    + N \left(Q_1 Q_2^2 + \frac{16}{3} Q_1^3\right)
     + \frac{B}{2 Q_2^2},
}
\end{array}
\right.
\end{eqnarray}
The confluence in this case is
\begin{eqnarray}
& &
\hbox{1:12:16 }\to\hbox{ KK}
\left\lbrace
\begin{array}{ll}
\displaystyle{
t=\varepsilon T,\
q_1=\varepsilon^{-1} + Q_1,\qquad
q_2=Q_2,\qquad
n=-\frac{4}{3}\varepsilon^{-1} N,\
}
\\
\displaystyle{
\alpha=\varepsilon^{-7}
 \left(-\frac{4}{3} N + 4 \Omega \varepsilon \right),\
\quad \beta=\varepsilon^{-2} B,\
}
\\
\displaystyle{
\omega=\varepsilon^{-3} \left(- 4 N + 4 \Omega \varepsilon \right),\ 
\qquad h=\varepsilon^{-5}
 \left(- 2 N + 4 \Omega \varepsilon + H \varepsilon^3\right),\
 \varepsilon \to 0,
}
\end{array}
\right.
\nonumber
\end{eqnarray}
and the two quartic parameters $(\alpha,\omega)$ coalesce to the single
cubic parameter $\Omega$.
\bigskip

We have checked that
all the generic cubic cases can be obtained by confluence from
at least one quartic case,
as indicated in the following list~:

\begin{eqnarray}
& &
\left\lbrace
\begin{array}{ll}
\displaystyle{
\hbox{HH4 1:2:1 } \to \hbox{ HH3 KdV5},\
}
\\
\noalign{\bigskip}
\displaystyle{
\hbox{HH4 1:6:8 }
 \to \hbox{ HH3 KK},\
}
\\
\noalign{\bigskip}
\displaystyle{
\hbox{HH4 1:6:8 }
 \to \hbox{ HH3 KdV5},\
}
\\
\noalign{\bigskip}
\displaystyle{
\hbox{HH4 1:12:16 } \to \hbox{ HH3 KK}.
}
\\
\noalign{\bigskip}
\displaystyle{
\hbox{HH4 1:12:16 } \to \hbox{ HH3 SK}.
}
\end{array}
\right.
\end{eqnarray}
\bigskip

Since these confluences are not invertible and always go from quartic to cubic,
they are unfortunately of no help to integrate the missing cases,
which are all quartic.
In section \ref{sectionBirational},
we present another class of transformations, these one invertible,
between some of the seven cases,
which indeed helps to integrate the missing cases.

\section     {Equivalent fourth order ODEs}
\label{sectionFourthOrderODEs}
\indent

The Painlev\'e school has ``classified''
(i.e, enumerated the integrable equations and integrated them)
several types of ODEs
(e.g.,~second order first degree, third order first degree of the polynomial type,
etc),
but no four-dimensional first order differential system such as the
Hamilton equations
\begin{eqnarray}
& &
\frac{\D q_j}{\D t}=p_j,\
\frac{\D p_j}{\D t}= - \frac{\partial V}{\partial q_j},\ j=1,2
\end{eqnarray}
has ever been classified.
However, some types of fourth order ODEs have been classified,
in particular the polynomial class
(\cite{ChazyThese,BureauMII,Cos2000a,Cos2000c})
\begin{eqnarray}
& & u''''=P(u''',u'',u',u,x),
\label{eqODE4PolynomialClass}
\end{eqnarray}
in which $P$ is polynomial in $u''',u'',u',u$ and analytic in $x$.
Therefore,
if one succeeds,
by elimination of either $q_1$ or $q_2$ (or another combination)
between the two Hamilton equations
and the equation $H=E$ expressing the conservation of the energy,
to build a fourth order ODE in the class
(\ref{eqODE4PolynomialClass}),
and if this ODE is \textit{equivalent} to the original system,
then the question is settled.

In the cubic case, the two Hamilton equations
\begin{eqnarray}
& & q_1'' + \omega_1 q_1 - \beta q_1^2 + \alpha q_2^2 = 0,
\label{eqHH1}
\\
\noalign{\medskip}
& & q_2'' + \omega_2 q_2 + 2 \alpha q_1 q_2 - \gamma q_2^{-3} = 0,
\label{eqHH2}
\end{eqnarray}
together with $H-E=0$, see (\ref{eqHH0}),
are indeed equivalent (\cite{Fordy1991})
to the single fourth order first degree ODE for $q_1(t)$,
\begin{eqnarray}
& &
 q_1'''' + (8 \alpha - 2 \beta) q_1 q_1''
 - 2 (\alpha + \beta) q_1'^2
 - \frac{20}{3} \alpha \beta q_1^3
\nonumber
\\
& &\qquad\qquad\qquad 
  +(\omega_1 + 4 \omega_2) q_1''
  + (6 \alpha \omega_1 - 4 \beta \omega_2) q_1^2 + 4 \omega_1 \omega_2 q_1
   + 4 \alpha E
=0.
\label{eqHH3ODE4}
\end{eqnarray}
The equivalence results from the conservation of the number of parameters
between the system (\ref{eqHH1})--(\ref{eqHH2}) and the single equation
(\ref{eqHH3ODE4}),
since the coefficient $\gamma$ of the nonpolynomial term $q_2^{-2}$
has been replaced by the constant value $E$ of the Hamiltonian $H$.
The results of the classification \cite{Cos2000a}
enumerate as expected only three Painlev\'e-integrable such equations
and they provide their general solution
(for the first time in the SK and KK cases).
\bigskip

In the quartic case,
the similar fourth order equation
is built by eliminating $Q_2$ and ${Q_1'''}^2$ between the
two Hamilton equations,
\begin{eqnarray}
& & Q_1''+\Omega_1 Q_1 + 4 C Q_1^3 + 2 B Q_1 Q_2^2 - \alpha Q_1^{-3} + \gamma=0,
\label{eqHH41}
\\
\noalign{\medskip}
& & Q_2''+\Omega_2 Q_2 + 4 A Q_2^3 + 2 B Q_2 Q_1^2 - \beta  Q_2^{-3}=0,
\label{eqHH42}
\end{eqnarray}
and the Hamiltonian (\ref{eqHH40}), which results in
\begin{eqnarray}
& &
-Q_1''''
+ 2 \frac{Q_1' Q_1'''}{Q_1}
+\left(1 + 6 \frac{A}{B}\right) \frac{{Q_1''}^2}{Q_1}
-2 \frac{{Q_1'}^2 Q_1''}{Q_1^2}
\nonumber \\ & & \phantom{12345}
+8 \left(6 \frac{A C}{B} - B - C\right) Q_1^2 Q_1''
+4(B - 2 C) Q_1 {Q_1'}^2
+24 C \left(4 \frac{A C}{B} - B\right) Q_1^5
\nonumber \\ & & \phantom{12345}
+\left\lbrack
12 \frac{A}{B} \omega_1 - 4 \omega_2
+\left(1 + 12 \frac{A}{B}\right) \frac{\gamma}{Q_1}
- 4 \left(1+3 \frac{A}{B}\right) \frac{\alpha}{Q_1^4}
\right\rbrack Q_1''
\\ & & \phantom{12345}
+ 6 \frac{A}{B} \frac{\alpha^2}{Q_1^7}
+ 20 \frac{\alpha} {Q_1^5} {Q_1'}^2
-12  \frac{A}{B} \frac{\gamma \alpha}{Q_1^4}
+4 \left(3 \frac{A}{B} \omega_1 - \omega_2 \right)
   \left(\gamma - \frac{\alpha}{Q_1^3}\right)
-2 \gamma \frac{{Q_1'}^2}{Q_1^2}
\nonumber \\ & & \phantom{12345}
+ 6 \left(\frac{A}{B} \gamma^2 + 2 B \alpha -8 \frac{A C}{B} \alpha\right)
     \frac{1}{Q_1}
+ \left(6 \frac{A}{B} \omega_1^2 -4 \omega_1 \omega_2 -8 B E\right) Q_1
\nonumber \\ & & \phantom{12345}
+ 48 \frac{A C}{B} \gamma Q_1^2
+ 4 \left(12  \frac{A C}{B} - B - 4 C \right) \omega_1 Q_1^3=0.
\nonumber\label{eqHH4odeq1}
\end{eqnarray}
\bigskip

The equivalence with the Hamilton equations results from the dependence
on $E$ but not on $\beta$.
However,
this type of fourth order first degree ODEs has not yet been classified,
and this would be quite useful to do so,
in order to check that no Painlev\'e-integrable case has been omitted
when performing the Painlev\'e test on the coupled system made of the two
Hamilton equations.

\section     {Birational transformations between the quartic cases
 and integrated equations}
\label{sectionBirational}
\indent

Between Hamiltonians with one degree of freedom
such as $H=p^2/2 + a q^2 + b q^3 + q^4$ and $H=p^2/2 + A q^2 + q^3$,
there exist invertible transformations which allow one to carry out the solution
from one case to the other.
These are the well known homographies between
the Jacobi and the Weierstrass elliptic functions.
In the present case of two degrees of freedom,
the simplest example of such a transformation is (\cite[Eq.~(7.14)]{CMVCalogero})

\begin{eqnarray}
& &
\left\lbrace
\begin{array}{ll}
\displaystyle{
Q_1^2 + Q_2^2 + \frac{\Omega_1+\Omega_2}{5}
= \alpha q_1  + \frac{\omega_1+4\omega_2}{20},\
}
\\
\noalign{\bigskip}
\displaystyle{
(\Omega_1-\Omega_2)(Q_1^2 - Q_2^2) =
\frac{\alpha^2}{2} q_2^2
- \frac{4 \omega_1 + 26 \omega_2}{5} \alpha q_1
- \frac{(\omega_1+4 \omega_2)^2}{100}
+2 E,
}
\\
\noalign{\bigskip}
\displaystyle{
\Omega_1=  \omega_1,\qquad \qquad
\Omega_2=4 \omega_2,\
}
\end{array}
\right.
\label{eqFrom_121_to_KdV5}
\end{eqnarray}
\medskip

\noindent between the quartic 1:2:1 case $H(Q_j,P_j,\Omega_1,\Omega_2,A,B)$
and the cubic KdV5 case $H(q_j,p_j,\omega_1,\omega_2,\alpha,\gamma)$.
Its action on the hyperelliptic curves is just a translation.

An attempt to find transformations between the other quartic cases
and any cubic case which would be as simple as (\ref{eqFrom_121_to_KdV5})
has been unsuccessful for the moment.
\smallskip

However,
it is possible to obtain a birational transformation (\cite{CMVCalogero})
between every remaining quartic case (1:6:1, 1:6:8, 1:12:16)
and some classified fourth order ODE of the type (\ref{eqODE4PolynomialClass}).
Indeed,
for each of the seven cases,
the two Hamilton equations are equivalent
(\cite{Fordy1991,FK1983,BakerThesis})
to the traveling wave reduction of a soliton system made
either of a single PDE (HH3) or of two coupled PDEs (HH4),
most of them appearing in lists established from group theory
(\cite{DS1981}).
Among the various soliton equations which are equivalent to them
\textit{via} a B\"acklund transformation,
some of them admit a traveling wave reduction to a classified ODE.
This property defines a path (\cite{MV2003,VMC2004a}) which
starts from one of the three remaining HH4 cases,
goes up to a soliton system of two coupled 1+1-dimensional PDEs
admitting a reduction to the considered case,
then goes \textit{via} a B\"acklund transformation
to another equivalent 1+1-dim PDE system,
finally goes down by reduction to an already integrated
ODE or system of ODEs.

\subsection  {Integration of the 1:6:1 and 1:6:8 cases with the a-F-VI equation}
\label{sectionIntegrationQuartic1:6:1and1:6:8}
\quad \smallskip

In this section, the integration is performed \textit{via} a birational
transformation to the autonomous F-VI equation (a-F-VI)
in the classification of Cosgrove (\cite{Cos2000a}):
\begin{eqnarray}
& &
\hbox{a-F-VI}:\
y''''=18 y y'' + 9 {y'}^2 - 24 y^3
+ \alpha_{\rm VI} y^2 + \frac{\alpha_{\rm VI}^2}{9} y
+ \kappa_{\rm VI} t + \beta_{\rm VI},\
\kappa_{\rm VI}=0.
\label{eqCosgroveFVI}
\end{eqnarray}

The two considered Hamiltonians,
with their second constant of the motion, are the following,
\begin{eqnarray}
& &
1:6:1
\left\lbrace
\begin{array}{ll}
\displaystyle{
H =
 \frac{1}{2}(P_1^2+P_2^2)
+\frac{\Omega}{2}(Q_1^2+Q_2^2)
\qquad\qquad
}
\\
\noalign{\medskip}
\hfill\displaystyle{\displaystyle
\phantom{1234}\qquad\qquad 
-\frac{1}{32} (Q_1^4+ 6 Q_1^2 Q_2^2 + Q_2^4)
 -\frac{1}{2}\Big(\frac{\kappa_1^2}{Q_1^2}+\frac{\kappa_2^2}{Q_2^2}\Big)
=E,
}
\\
\noalign{\bigskip}
\displaystyle{
K =
\left(
P_1 P_2 + Q_1 Q_2 \Big(-\frac{Q_1^2+Q_2^2}{8}+\Omega \Big)
\right)^2
}
\\
\noalign{\medskip}
\hfill\qquad\qquad\displaystyle{
\phantom{1234}
- P_2^2 \frac{\kappa_1^2}{Q_1^2}
- P_1^2 \frac{\kappa_2^2}{Q_2^2}
+\frac{1}{4}\left(\kappa_1^2 Q_2^2 + \kappa_2^2 Q_1^2 \right)
+\frac{\kappa_1^2 \kappa_2^2}{Q_1^2 Q_2^2},
}
\end{array}
\right.
 \label{eqHH40161}
\end{eqnarray}
and
\begin{eqnarray}
& &
1:6:8
\left\lbrace
\begin{array}{ll}
\displaystyle{
H =
 \frac{1}{2}(p_1^2+p_2^2)
+\frac{\omega}{2}(4 q_1^2+q_2^2)
}
\\
\noalign{\medskip}
\hfill\qquad\qquad\displaystyle{
\phantom{1234}
-\frac{1}{16} (8 q_1^4+ 6 q_1^2 q_2^2 + q_2^4)
- \gamma q_1 +\frac{\beta}{2 q_2^2}
=E,
}
\\
\noalign{\bigskip}
\displaystyle{
K =
\left(
p_2^2-\frac{q_2^2}{16}(2 q_2^2+4 q_1^2+\omega)
     +\frac{\beta}{q_2^2}
\right)^2
}
\\
\noalign{\medskip}
\displaystyle
\phantom{1234}
\hfill\ -\frac{1}{4}q_2^2(q_2 p_1 - 2 q_1 p_2)^2
+\gamma
\Big(
-2 \gamma q_2^2
-4 q_2 p_1 p_2\qquad\\
\noalign{\medskip}
\displaystyle
\hfill\qquad\qquad +\frac{1}{2} q_1 q_2^4
+ q_1^3 q_2^2
+4 q_1 p_2^2
-4 \omega q_1 q_2^2
+ 4 q_1 \frac{\beta}{q_2^2}
\Big).
\end{array}
\right.
\label{eqHH40168}
\end{eqnarray}
\bigskip

There exists a canonical transformation (\cite{BakerThesis})
between these two cases, mapping the constants as follows:
\begin{eqnarray}
\hspace{1cm} E_{1:6:8}=E_{1:6:1},\ \ 
K_{1:6:8}=K_{1:6:1},\ \ 
\omega=\Omega,\ \ 
\gamma=\frac{\kappa_1+\kappa_2}{2},\ \ 
\beta=- (\kappa_1-\kappa_2)^2.
\label{eqCT161to168constants}
\end{eqnarray}
Therefore, one only needs to integrate either case.
\bigskip

The path to an integrated ODE comprises the following three segments.

The coordinate $q_1(t)$ of the 1:6:8 case can be identified
(\cite{BEF1995b,BakerThesis})
to the component $F$ of
the traveling wave reduction
$f(x,\tau)=F(x-c \tau), g(x,\tau)=G(x-c \tau)$
of a soliton system of two coupled KdV-like equations
(\cKdV\ system)
denoted \cKdV${}_1$
(\cite{BEF1995b,BakerThesis})
\begin{eqnarray}
{\hskip -10.0 truemm}
& &
\left\lbrace
\begin{array}{ll}
\displaystyle{
f_\tau+ \left(f_{xx}+\frac{3}{2} f f_{x}-\frac{1}{2}f^3+3 f g\right)_x=0,
}
\\
\noalign{\bigskip}
\displaystyle{
- 2 g_\tau+ g_{xxx}+ 6 g g_x+3 f g_{xx}+6 gf_{xx}+9 f_x g_x-3 f^2 g_x
\qquad\qquad\qquad
}
\\
\noalign{\medskip}
\displaystyle\hfill
\phantom{xxxxx}+\frac{3}{2} f_{xxxx} +\frac{3}{2} f f_{xxx}+9 f_x f_{xx}
               -3 f^2 f_{xx}-3 f f_x^2=0,
\end{array}
\right.
\label{eq:cKdV1}
\end{eqnarray}
with the identification
\begin{eqnarray}
{\hskip -10.0 truemm}
& &
\left\lbrace
\begin{array}{ll}
\displaystyle{
q_1=F,\
q_2^2=-2\left(F'+F^2 + 2 G -2 \omega\right),\
}
\\
\noalign{\bigskip}
\displaystyle{
c=-\omega,\
K_1=\gamma,\
K_2=E,
}
\end{array}
\right.
\label{eqcKdV1r_to_HH4168}
\end{eqnarray}
in which $K_1$ and $K_2$ are two constants of integration.

There exists a B\"acklund transformation between this soliton system
and another one of \cKdV\ type, denoted bi-SH system (\cite{DS1981}):
\begin{eqnarray}
& &
\left\lbrace
\begin{array}{ll}
\displaystyle{
- 2 u_\tau+ \left(u_{xx} + u^2 + 6 v\right)_x = 0,
}
\\
\noalign{\bigskip}
\displaystyle{
v_\tau+ v_{xxx} + u v_x = 0.
}
\end{array}
\right.
\label{eq:systemHSII}
\end{eqnarray}
This B\"acklund transformation  is defined by the Miura transformation (\cite{MV2003})
\begin{eqnarray}
{\hskip -10.0 truemm}
& &
\left\lbrace
\begin{array}{ll}
\displaystyle{
u=\frac{3}{2} \left(2 g -f_x-f^2\right),
}
\\
\noalign{\bigskip}
\displaystyle{
v=\frac{3}{4}
\left(
 2 f_{xxx}
+4 f f_{xx}
+8 g f_x
+4 f g_x
+3 f_x^2
-2 f^2 f_x
-  f^4
+4 g f^2\right).
}
\end{array}
\right.
\end{eqnarray}

Finally,
the traveling wave reduction
$$\left\lbrace\begin{array}{ccl}u(x,\tau)&=&U(x-c \tau),\\
v(x,\tau)&=&V(x-c \tau)
\end{array}\right.$$
can be identified (\cite{VMC2004a})
to the autonomous F-VI equation (a-F-VI)
(\ref{eqCosgroveFVI}),
whose general solution is meromorphic,
expressed with genus two hyperelliptic functions (\cite[Eq.~(7.26)]{Cos2000a}).
The identification is
\begin{eqnarray}
{\hskip -10.0 truemm}
& &
\left\lbrace
\begin{array}{ll}
\displaystyle{
U=- 6 \left(y+\frac{c}{18}\right),\
}
\\
\noalign{\bigskip}
\displaystyle{
V=y'' -6 y^2 + \frac{4}{3} c y
+\frac{16}{27} c^2 -\frac{K_A}{2},
}
\\
\noalign{\bigskip}
\displaystyle{
\alpha_{\rm VI}=-4 c,\
\beta_{\rm VI}=K_B-2 c K_A + \frac{512}{243} c^3,
}
\end{array}
\right.
\label{eqFVI_to_biSHr}
\end{eqnarray}
in which $K_A,K_B$ are two constants of integration.
\bigskip

In order to perform the integration of both the 1:6:1 and the 1:6:8 cases,
it is sufficient
to express $(F,G)$ rationally in terms of $(U,V,U',V')$.
The result is
\begin{eqnarray}
& &
\left\lbrace
\begin{array}{ccl}
F&=&\displaystyle \frac{W'}{2 W}
+ \frac{K_1}{24 W}
\left[ -3 {U'}^2
\right.\qquad\qquad\qquad\qquad
\\
\noalign{\medskip}
&&\displaystyle\left.
\hfill\qquad\qquad
-2 (U-3 c) \left(12 V + (U + 3 c)^2\right)
+36 K_B
-54 K_1^2
\right],
\\
\noalign{\bigskip}
G&=&\displaystyle
\frac{U}{3} + \frac{1}{8 W}
\left[
(2 V + 3 K_2) \left(2 V'' + K_1 U' -3 K_1^2 \right)
\right.
\\
\noalign{\bigskip}
&&\left.\hfill\qquad\quad
\phantom{1234567890123}
-2 (U - 3 c)\left(2 K_1 V' + K_1^2 (U + 3 c)\right)
\right],
\\
\noalign{\bigskip}
W&=&\displaystyle
\left(V + \frac{3}{2} K_2\right)^2 + \frac{3}{2} K_1^2 (U - 3 c),
\\
\noalign{\bigskip}
K_A&=&K_2.
\end{array}
\right.
\label{eqbiSHr_to_cKdV1r}
\end{eqnarray}

Making the product of the successive transformations
(\ref{eqcKdV1r_to_HH4168}),
(\ref{eqbiSHr_to_cKdV1r}),
(\ref{eqFVI_to_biSHr}),
one obtains
a meromorphic general solution
for $q_1,q_2^2$: 
\bigskip

\begin{eqnarray}
\left\lbrace
\begin{array}{rl}
q_1&=\displaystyle\frac{W'}{2 W}
+ \frac{\gamma}{W}
\left[9 j -3 \left(y + \frac{4}{9}\omega\right) (h+E)
      - \frac{9}{4} \gamma^2\right],
\\
\noalign{\bigskip}
q_2^2&=\displaystyle-16 \left(y - \frac{5}{9} \omega\right)
+\frac{1}{W} \Big[ 12 \left(y' + \frac{\gamma}{2}\right)^2
 -48 y^3 - 16 \omega y^2
\\
\noalign{\medskip}
&\displaystyle
\phantom{123}
 +\left(24 E +\frac{128}{9}\omega^2\right) y
 + \frac{1280}{243} \omega^3
- \frac{40}{3} \omega E + \frac{3}{4} \beta
\Big.
\\ 
\noalign{\medskip}
&\displaystyle{\phantom{12345678}
\Big.
-24 \gamma \left(y - \frac{5}{9} \omega\right) h'
-144 \gamma^2 \left(y - \frac{5}{9} \omega\right)^2
\Big],
}
\\
\noalign{\bigskip}
W&=\displaystyle (h+E)^2 -9 \gamma^2 \left(y - \frac{5}{9} \omega\right),
\\
\noalign{\bigskip}
\alpha_{\rm VI}&=\displaystyle 4 \omega,\qquad \qquad \qquad \quad
\beta_{\rm VI}=\displaystyle\frac{3}{4} \gamma^2 + 2 \omega E
               - \frac{3}{16} \beta-\frac{512}{243} \omega^3,
\\
\noalign{\bigskip}
K_{1,{\rm VI}}&=\displaystyle\frac{3}{32}   K - \frac{1}{2} E^2,\qquad\ \ 
K_{2,{\rm VI}}=\frac{3}{32} E K - \frac{1}{3} E^3+\frac{9}{64} \beta \gamma^2,
\\
\noalign{\bigskip}
K_1&=\displaystyle\gamma, \qquad
K_2=E,\qquad
K_A=E, \qquad
K_B=- \frac{3}{16} \beta + \frac{3}{4} \gamma^2,
\end{array}
\right.
\label{eqFVI_to_HH4168}
\end{eqnarray}
\bigskip

\noindent 
in which $h$ and $j$ are the convenient auxiliary variables
(\cite[Eqs.~(7.4)--(7.5)]{Cos2000a})
\begin{eqnarray}
{\hskip -10.0 truemm}
& &
\left\lbrace
\begin{array}{ll}
y&=\displaystyle\frac{Q(s_1,s_2) + \sqrt{Q(s_1) Q(s_2)}}
       {2\left(\sqrt{s_1^2-C_{\rm VI}}+\sqrt{s_2^2-C_{\rm VI}} \right)^2}
 + \frac{5}{36} \alpha_{\rm VI},\
\\
\noalign{\bigskip}
h&=\displaystyle-\frac{3}{4} E_{\rm VI}
 \frac{s_1 s_2 + C_{\rm VI} + \sqrt{(s_1^2-C_{\rm VI})(s_2^2-C_{\rm VI})}}
      {s_1+s_2}
 - \frac{F_{\rm VI}}{2},\
\\
\noalign{\bigskip}
j&=\displaystyle\frac{1}{6} (2 h + F_{\rm VI})
 \left\lbrace
y + \frac{\alpha_{\rm VI}}{9} - \frac{E_{\rm VI}}{4(s_1+s_2)}
\right\rbrace .
\end{array}
\right.
\label{eqFVI_General_solution}
\end{eqnarray}
\bigskip

\noindent In the above,
the variables $s_1,s_2$ are defined by the
hyperelliptic system \cite{Cos2000a}
\begin{eqnarray}
& &
\left\lbrace
\begin{array}{ll}
\displaystyle{
(s_1-s_2)s_1'=\sqrt{P(s_1)},\qquad\quad
(s_2-s_1)s_2'=\sqrt{P(s_2)},\
}
\\
\noalign{\bigskip}
\displaystyle{
P(s)=(s^2-C_{\rm VI}) Q(s),
}
\\
\noalign{\bigskip}
\displaystyle{
Q(s,t)=(s^2-C_{\rm VI})(t^2-C_{\rm VI})
- \frac{\alpha_{\rm VI}}{2} (s^2+t^2-2 C_{\rm VI})
        +\frac{E_{\rm VI}}{2}(s+t)+F_{\rm VI},
}
\\
\noalign{\bigskip}
\displaystyle{
Q(s)=Q(s,s).
}
\end{array}
\right.
\label{eqHyperellipticGenusTwoSystem}
\label{eqCosgroveFVIsextic}
\end{eqnarray}
\bigskip

\noindent Despite their square roots,
the symmetric expressions in (\ref{eqFVI_General_solution})
are nevertheless meromorphic (\cite{FarkasKra,Mumford}).
\bigskip

The completeness of both the 1:6:1 and 1:6:8 Hamiltonians results from
the completeness of the a-F-VI ODE and the following counting.
The 1:6:8 depends on the parameters $(\omega,\beta,\gamma,E,K)$,
the a-F-VI ODE and its hyperelliptic system depend on the same number of parameters
$(\alpha,\beta,C,E,F)_{\rm VI}$,
and these two sets of five parameters are linked by exactly five
algebraic relations
(\cite[Eqs.~(7.9)-(7.12)]{Cos2000a}):
\begin{eqnarray}
{\hskip -10.0 truemm}
& &
\left\lbrace
\begin{array}{l}
\alpha_{\rm VI}\displaystyle=4 \omega,
\\
\noalign{\bigskip}
\beta_{\rm VI}\displaystyle=\frac{3}{4} \gamma^2 + 2 \omega E
               - \frac{3}{16} \beta-\frac{512}{243} \omega^3,
\\
\noalign{\bigskip}
E_{\rm VI}^2\displaystyle=-\frac{16}{3}\omega(F_{\rm VI}-2 E)-\beta  
+ 4 \gamma^2,
\\
\noalign{\bigskip}
C_{\rm VI} E_{\rm VI}^2\displaystyle=\frac{4}{3}(F_{\rm VI}^2-4 E^2) +K,
\\
\noalign{\bigskip}
(F_{\rm VI}-2 E)^2 (F_{\rm VI}+4 E) + \frac{9 K}{4} (F_{\rm VI}-2 E)
- \frac{27}{4} \beta \gamma^2=0.
\end{array}
\right.
\label{eqFVI_to_HH4168_const}
\end{eqnarray}
The algebraic nature
(instead of rational like in the 1:2:1 case and the three cubic cases)
of these dependence relations could explain the difficulty
to separate the variables in the Hamilton-Jacobi equation.
In the nongeneric case $\beta \gamma=0$, i.e, $\kappa_1^2=\kappa_2^2$,
for which the separating variables are known (\cite{RRG}),
the coefficients $(\alpha,\beta,C,E^2,F)_{\rm VI}$
become rational functions of $(\omega,\beta,\gamma,E,K)$, see \cite{VMC2003}.
Since these separating variables have been obtained by the same method
as in the cubic SK-KK case,
it would be quite useful to remove the difficulty
which remains in the method based on Laurent series,
see Section \ref{sectionCasesWithSeparatingVariables}.

\subsection  {Integration of the 1:12:16 case by a birational transformation}
\label{sectionIntegrationQuartic1:12:16}
\quad\medskip

This is the only case for which the integration,
which can indeed be performed with the same results
(meromorphy of the general solution, completeness of the Hamiltonian)
is not satisfying.
Indeed,
the hyperelliptic system to which the 1:12:16 has been mapped
by a birational transformation \cite{CMVCalogero}
is essentially different from
the hyperelliptic system resulting from the separating variables \cite{V2003}
in the nongeneric case $\alpha \beta=0$ for which they are known.
Since the nongeneric subcase $\alpha=0$ belongs to the St\"ackel class
(two invariants quadratic in $p_1,p_2$),
for which the separating variables are unambiguous,
this indicates that some progress has still to be made.

The main remarkable feature of the 1:12:16 is the existence of a
twin system to which it is mapped by a canonical transformation
\cite{BakerThesis,BEF1995b}
which only differs by numerical coefficients
from the canonical transformation between the cubic SK and KK cases.
The two systems are the following ones:
\begin{eqnarray}
& &
\ 1:12:16
\left\lbrace
\begin{array}{ccl}
H& =&\displaystyle
\frac{1}{2}(P_1^2+P_2^2) + \frac{\Omega}{8} (4 Q_1^2+ Q_2^2)\qquad\qquad
\\
\noalign{\medskip}
&&\displaystyle\hfill \qquad
 - \frac{1}{32} (16 Q_1^4+ 12 Q_1^2 Q_2^2 + Q_2^4)
 -\frac{1}{2}\Big(\frac{\kappa_1^2}{Q_1^2}+ 
 \frac{4 \kappa_2^2}{Q_2^2}\Big)
=E,
\\
\noalign{\bigskip}
K &=&\displaystyle
\frac{1}{16}
\Big(8 (Q_2 P_1 - Q_1 P_2) P_2 - Q_1 Q_2^4 - 2 Q_1^3 Q_2^2
\\
\noalign{\bigskip}
&&\displaystyle\hfill\qquad
 + 2 \Omega Q_1 Q_2^2 + 32 Q_1 \frac{\kappa_2^2}{Q_2^2} \Big)^2
+ \kappa_1^2 \left(Q_2^4 - 4 \frac{Q_2^2 P_2^2}{Q_1^2}\right).
\end{array}
\right.
\label{eqHH401:12:16E}
\label{eqHH401:12:16K}
\end{eqnarray}
and
[this system is not the sum of a kinetic energy and a potential energy]
\begin{eqnarray}
&&
\ 5:9:4
\left\lbrace
\begin{array}{l}
H =\displaystyle
\frac{1}{2}\left(p_1^2+\left(p_2 - \frac{3}{2} q_1 q_2\right)^2\right)
-\frac{1}{8} (4 q_1^4 + 9 q_1^2 q_2^2 +5 q_2^4)
\\
\noalign{\bigskip}
\displaystyle \hfill\qquad+\frac{\omega}{2}(q_1^2+q_2^2)
-\kappa q_1
+\frac{\zeta}{2 q_2^2}
=E,
\\
\noalign{\bigskip}
K=\displaystyle\frac{1}{q_2^2}
\left(
  2 q_2^2 p_1
+ 2 q_1^2 q_2^2
-2 q_1 q_2 p_2
- q_2^4
 - 4 \kappa q_1
\right)^2 \bigg(
  2 q_2^2 p_1
\\
\noalign{\bigskip}
\displaystyle\hfill\qquad
+ 2 q_1^2 q_2^2
+ p_2^2 -4 q_1 q_2 p_2
- 2 q_2^4
 + \Omega q_2^2
 + 4\frac{\kappa^2}{q_2^2}
 + 8 \kappa q_1
 -4 \kappa \frac{p_2}{q_2}
\\
\noalign{\bigskip}
\displaystyle \hfill
+ 4 (\zeta + 4 \kappa^2)
\bigg)
\left(
\Big(-2 q_1 \frac{p_2}{q_2}+4 q_1^2+q_2^2+4 q_1 
\frac{\kappa}{q_2^2}\Big) p_1
\right.
\\
\noalign{\bigskip}
\displaystyle \hfill
 - \frac{1}{q_2^4} (q_1^2 q_2^2 + q_2^4 + 2 \kappa q_1)^2
+2 \frac{q_1^2}{q_2^2} \Big(p_2 - \frac{3}{2} q_1 q_2 \Big)^2
\\
\noalign{\bigskip}
\displaystyle \hfill
\left.
+ \frac{(q_1^2+q_2^2)^2}{2}
+q_1^2 \frac{\zeta}{q_2^4}
\right).
\end{array}
\right.
\label{eqHH405:9:4E}
\label{eqHH405:9:4K}
\end{eqnarray}

The canonical transformation maps the constants as follows:
\begin{eqnarray}
& &
E_{5:9:4}=E_{1:12:16},\ \ 
K_{5:9:4}=K_{1:12:16},\ \ 
\omega=\Omega,\ \ 
\kappa=\frac{\kappa_1 + \kappa_2}{2},\ \ 
\zeta=-(\kappa_1 - \kappa_2)^2.
\end{eqnarray}

The path to an integrated ODE is quite similar to that described in detail
in section \ref{sectionIntegrationQuartic1:6:1and1:6:8},
in particular it is also made
of three segments (\cite{BakerThesis,V2003,MV2003}).
The result is the following (\cite{CMVCalogero}):
\begin{eqnarray}
& &
Q_1,Q_2^2=\hbox{rational}(y,y',y'',y'''),
\end{eqnarray}
in which $y$ obeys the F-IV equation
in the classification of Cosgrove (\cite{Cos2000a}),
\medskip

\begin{eqnarray}
& \hbox{F-IV} &
\left\lbrace
\begin{array}{l}
\displaystyle
y''''=30 y y'' - 60 y^3 + \alpha_{\rm IV} y + \beta_{\rm IV},
\\
\noalign{\bigskip}
\displaystyle
\ y\ =\frac{1}{2}
 \left(s_1'+s_2' + s_1^2+s_1 s_2 + s_2^2 +A \right),
\\
\noalign{\bigskip}
\displaystyle
(s_1-s_2)s_1'=\sqrt{P(s_1)},\qquad 
(s_2-s_1)s_2'=\sqrt{P(s_2)},\
\\
\noalign{\bigskip}
\displaystyle
P(s)=(s^2+A)^3-\frac{\alpha_{\rm IV}}{3} 
(s^2+A) + B s + \frac{\beta_{\rm IV}}{3},
\\
\noalign{\bigskip}
\displaystyle
\KFIVone=\left(\frac{3 B}{4} \right)^2,\qquad \qquad
\KFIVtwo=-\frac{9 A B^2}{64}
\end{array}
\right.
\label{eqCosgroveFIV}
\end{eqnarray}
\medskip

\noindent with $(\KFIVone,\KFIVtwo)$ two polynomial first integrals of F-IV.
The general solution of this ODE is meromorphic,
expressed with genus two hyperelliptic functions (\cite{Cos2000a}).
This proves the PP for the 1:12:16.
\bigskip

In the two nongeneric cases $\kappa_1 \kappa_2=0$ where the separating
variables are known,
the hyperelliptic curve is
\begin{eqnarray}
& &
\kappa_1 \kappa_2=0:\
P(s)=s^6 - \omega s^3 + 2 E s^2 + \frac{K}{20} s
 + \kappa_1^2 + \kappa_2^2=0,
\label{eq1:12:16kappa1kappa20}
\end{eqnarray}
and it does not coincide in this case with the hyperelliptic curve
of F-IV.
Therefore, \hbox{F-IV} (as well as its birationally equivalent ODE F-III)
is not the good ODE to consider,
and it should be quite instructive to directly integrate the fourth order
equivalent ODE (\ref{eqHH4odeq1}) in that case.

\section{Conclusion and open problems}
\label{sectionOpen_problems}

All the time independent two-degree-of-freedom Hamiltonians
which possess the Painlev\'e property have a meromorphic general solution,
expressed with hyperelliptic functions of genus two.
Moreover,
all such Hamiltonians are complete in the Painlev\'e sense,
i.e, it is impossible to add any term to the Hamiltonian without ruining
the Painlev\'e property.
\bigskip

As to the remaining open problems,
depending on the center of interest, they are
\begin{enumerate}
\item
from the point of view of Hamiltonian theory,
one has to find the
separating variables in the three missing quartic cases.
This should be possible by the methods of Sklyanin
and van Moerbeke and Vanhaecke;

\item
from the point of view of the integration of differential equations,
the problem remains to enumerate all
the fourth order first degree differential equations in a given precise class
which possess the Painlev\'e property.

\end{enumerate}

Let us finally mention that the time dependent extension of these seven cases
has been studied in Refs.~\cite{Hone1998HH3,Hone2001HH4}.

\backmatter
\bibliographystyle{smfplain}
\nocite{*}
\bibliography{Angers} 

\providecommand{\bysame}{\leavevmode ---\ }
\providecommand{\og}{``}
\providecommand{\fg}{''}
\providecommand{\smfandname}{\&}
\providecommand{\smfedsname}{\'eds.}
\providecommand{\smfedname}{\'ed.}
\providecommand{\smfmastersthesisname}{M\'emoire}
\providecommand{\smfphdthesisname}{Th\`ese}
\begin{thebibliography}{10}

\bibitem{AMB}
{\scshape S.~Abenda, V.~Marinakis {\normalfont \smfandname} T.~Bountis} -- {\og
  {O}n the connection between hyperelliptic separability and {P}ainlev\'e
  integrability\fg}, \emph{J.~Phys.~A} (2001), no.~34, p.~3521--3539.

\bibitem{AvM1987}
{\scshape M.~Adler {\normalfont \smfandname} P.~van Moerbeke} --
  \emph{Completely integrable system -- {A} systematic approach},
  \textit{Perspective in Mathematics}, Academic press, New York, 1987?

\bibitem{AvM1989}
{\scshape M.~Adler {\normalfont \smfandname} P.~van Moerbeke} -- {\og The
  complex geometry of the {K}owalewski-{P}ainlev\'e analysis\fg},
  \emph{Invent.~Math.} (1989), no.~97, p.~3--51.

\bibitem{AP1983}
{\scshape A.~Ankiewicz {\normalfont \smfandname} C.~Pask} -- {\og The complete
  {W}hittaker theorem for two-dimensional integrable systems and its
  application\fg}, \emph{J.~Phys.~A} (1983), no.~16, p.~4203--4208.

\bibitem{ArnoldMechanics}
{\scshape V.~Arnol'd} -- \emph{Les m\'ethodes math\'ematiques de la m\'ecanique
  classique}, Nauka, Moscou, 1974, Mir, Moscou, 1976.

\bibitem{BakerThesis}
{\scshape S.~Baker} -- \emph{Squared eigenfunction representations of
  integrable hierarchies}, PhD Thesis, University of Leeds, Leeds, 1995.

\bibitem{BEF1995b}
{\scshape S.~Baker, V.~Z. Enol'skii {\normalfont \smfandname} A.~P. Fordy} --
  {\og Integrable quartic potentials and coupled {KdV} equations\fg},
  \emph{Phys.~Lett.~A} (1995), no.~201, p.~167--174.

\bibitem{BW1994}
{\scshape M.~B{\l}aszak {\normalfont \smfandname} S.~Rauch-Wojciechowski} --
  {\og A generalized {H}\'enon-{H}eiles system and related integrable {N}ewton
  equations\fg}, \emph{J.~Math.~Phys.} (1994), no.~35, p.~1693--1709.

\bibitem{BureauMII}
{\scshape F.~J. Bureau} -- {\og Differential equations with fixed critical
  points\fg}, \emph{Annali di Matematica pura ed applicata} (1964), no.~66,
  p.~1--116.

\bibitem{CTW}
{\scshape M.~T. Chang Y.~F. {\normalfont \smfandname} J.~Weiss} -- {\og
  Analytic structure of the {H}\'enon-{H}eiles {H}a\-miltonian in integrable
  and nonintegrable regimes\fg}, \emph{J.~Math.~Phys.} (1982), no.~23,
  p.~531--538.

\bibitem{ChazyThese}
{\scshape J.~Chazy} -- {\og Sur les \'equations diff\'erentielles du
  troisi\`eme ordre et d'ordre sup\'erieur dont l'int\'egrale g\'en\'erale a
  ses points critiques fixes\fg}, \emph{Acta Math.} (1911), no.~34,
  p.~317--385.

\bibitem{Cargese1996Conte}
{\scshape R.~Conte} -- {\og The {P}ainlev\'e approach to nonlinear ordinary
  differential equations\fg}, in \emph{The {P}ainlev\'e property, one century
  later}, CRM Series in Mathematical Physics (Springer), New York, 1999,
  p.~77--180.

\bibitem{CFP1993}
{\scshape R.~Conte, A.~P. Fordy {\normalfont \smfandname} A.~Pickering} -- {\og
  A perturbative {P}ainlev\'e approach to nonlinear differential equations\fg},
  \emph{Physica D} (1993), no.~69, p.~33--58.

\bibitem{CMVGallipoli2004}
{\scshape R.~Conte, M.~Musette {\normalfont \smfandname} C.~Verhoeven} -- {\og
  Completeness of the cubic and quartic {H}\'enon-{H}eiles {H}amiltonians\fg},
  \emph{Theor.~Math.~Phys.} (2005), no.~144, p.~888--898.

\bibitem{CMVCalogero}
\bysame , {\og Explicit integration of the {H}\'enon-{H}eiles
  {H}amiltonians\fg}, \emph{J.~Nonlinear Mathematical Physics} (2005), no.~12
  Supp.~1, p.~212--227.

\bibitem{Cos2000c}
{\scshape C.~M. Cosgrove} -- {\og Higher order {P}ainlev\'e equations in the
  polynomial class, i. {B}ureau symbol $p1$\fg}, \emph{preprint, University of
  Sydney} (2000), no.~2000--6, p.~1--113.

\bibitem{Cos2000a}
\bysame , {\og Higher order {P}ainlev\'e equations in the polynomial class, i.
  {B}ureau symbol $p2$\fg}, \emph{Stud.~Appl.~Math.} (2000), no.~104, p.~1--65.

\bibitem{Drach1919KdV}
{\scshape J.~Drach} -- {\og Sur l'int\'egration par quadratures de l'\'equation
  $\hbox{d}^2 y/ \hbox{d} x^2 = \lbrack\varphi(x) + h\rbrack y,$\fg},
  \emph{C.~R.~Acad.~Sc.~Paris} (1919), no.~168, p.~337--340.

\bibitem{DS1981}
{\scshape V.~G. Drinfel'd {\normalfont \smfandname} V.~V. Sokolov} -- {\og
  Equations of {K}orteweg-de {V}ries type and simple {L}ie algebras\fg},
  \emph{Soviet Math.~Dokl.} (1981), no.~23, p.~457--462.

\bibitem{FarkasKra}
{\scshape N.~Farkas {\normalfont \smfandname} E.~Kra} -- \emph{Riemann
  surfaces}, Springer, Berlin, 1980.

\bibitem{Fordy1991}
{\scshape A.~P. Fordy} -- {\og The {H}\'enon-{H}eiles system revisited\fg},
  \emph{Physica D} (1991), no.~52, p.~204--210.

\bibitem{FK1983}
{\scshape A.~P. Fordy {\normalfont \smfandname} P.~P. Kulish} -- {\og Nonlinear
  {S}chr\"odinger equations and simple {L}ie algebras\fg},
  \emph{Commun.~Math.~Phys.} (1983), no.~89, p.~427--443.

\bibitem{GDR1983}
{\scshape B.~Grammaticos, B.~Dorizzi {\normalfont \smfandname} A.~Ramani} --
  {\og Integrability of {H}amiltonians with third- and fourth-degree polynomial
  potentials\fg}, \emph{J.~Math.~Phys.} (1983), no.~24, p.~2289--2295.

\bibitem{HH}
{\scshape M.~H\'enon {\normalfont \smfandname} C.~Heiles} -- {\og The
  applicability of the third integral of motion: some numerical
  experiments\fg}, \emph{Astron.~J.} (1964), no.~69, p.~73--79.

\bibitem{H1984}
{\scshape J.~Hietarinta} -- {\og Classical versus quantum integrability\fg},
  \emph{J.~Math.~Phys.} (1984), no.~25, p.~1833--1840.

\bibitem{H1987}
\bysame , {\og Direct method for the search of the second invariant\fg},
  \emph{Phys.~Rep.} (1987), no.~147, p.~87--154.

\bibitem{Hone1998HH3}
{\scshape A.~N.~W. Hone} -- {\og Nonautonomous {H}\'enon-{H}eiles systems\fg},
  \emph{Physica D} (1998), no.~118, p.~1--16.

\bibitem{Hone2001HH4}
\bysame , {\og Coupled {P}ainlev\'e systems and quartic potentials\fg},
  \emph{J.~Phys.~A} (2001), no.~34, p.~2235--2245.

\bibitem{JM1983}
{\scshape M.~Jimbo {\normalfont \smfandname} T.~Miwa} -- {\og Solitons and
  infinite dimensional {L}ie algebras\fg}, \emph{Publ.~RIMS, Kyoto} (1983),
  no.~19, p.~943--1001.

\bibitem{Mumford}
{\scshape D.~Mumford} -- \emph{Tata lectures on theta, {II}}, Birkh\"auser,
  Basel, 1983.

\bibitem{MV2003}
{\scshape M.~Musette {\normalfont \smfandname} C.~Verhoeven} -- {\og On {CKP}
  and {BKP} equations related to the generalized quartic {H}\'enon-{H}eiles
  {H}amiltonian\fg}, \emph{Theor.~Math.~Phys.} (2003), no.~137, p.~1561--1573.

\bibitem{RDG1982}
{\scshape A.~Ramani, B.~Dorizzi {\normalfont \smfandname} B.~Grammaticos} --
  {\og Painlev\'e conjecture revisited\fg}, \emph{Phys.~Rev.~Lett.} (1982),
  no.~49, p.~1539--1541.

\bibitem{RGC}
{\scshape V.~Ravoson, L.~Gavrilov {\normalfont \smfandname} R.~Caboz} -- {\og
  Separability and {L}ax pairs for {H}\'enon-{H}eiles system\fg},
  \emph{J.~Math.~Phys.} (1993), no.~34, p.~2385--2393.

\bibitem{RRG}
{\scshape V.~Ravoson, A.~Ramani {\normalfont \smfandname} B.~Grammaticos} --
  {\og Generalized separability for a {H}amiltonian with nonseparable quartic
  potential\fg}, \emph{Phys.~Letters A} (1994), no.~191, p.~91--95.

\bibitem{Rom1995b}
{\scshape F.~J. Romeiras} -- {\og A note on integrable two-degrees-of-freedom
  hamiltonian systems with a second integral quartic in the momenta\fg},
  \emph{J.~Phys.~A} (1995), no.~28, p.~5633--5642.

\bibitem{SEL}
{\scshape M.~Salerno, V.~Z. Enol'skii {\normalfont \smfandname} D.~V. Leykin}
  -- {\og Canonical transformation between integrable {H}\'enon-{H}eiles
  systems\fg}, \emph{Phys.~Rev.~E} (1994), no.~49, p.~5897--5899.

\bibitem{Sklyanin1995}
{\scshape E.~K. Sklyanin} -- {\og Separation of variables -- {N}ew trends
  --\fg}, \emph{Prog.~Theor.~Phys.~Suppl.} (1995), p.~35--60.

\bibitem{Staeckel1893a}
{\scshape P.~St\"ackel} -- {\og Sur une classe de probl\`emes de
  {D}ynamique\fg}, \emph{C.~R.~Acad.~Sc.~Paris} (1893), no.~116, p.~485--487.

\bibitem{Staeckel1893b}
\bysame , {\og Sur une classe de probl\`emes de {D}ynamique, qui se r\'eduisent
  \`a des quadratures\fg}, \emph{C.~R.~Acad.~Sc.~Paris} (1893), no.~116,
  p.~1284--1286.

\bibitem{VanhaeckeLNM}
{\scshape P.~Vanhaecke} -- \emph{Integrable systems in the realm of algebraic
  geometry}, Lecture notes in mathematics, vol. 1638, Springer, Berlin, 1996.

\bibitem{V2003}
{\scshape C.~Verhoeven} -- \emph{Integration of {H}amiltonian systems of
  {H}\'enon-{H}eiles type and their associated soliton equations}, PhD Thesis,
  Vrije Universiteit Brussel, Bruxelles, 2003.

\bibitem{VMC2002a}
{\scshape C.~Verhoeven, M.~Musette {\normalfont \smfandname} R.~Conte} -- {\og
  Integration of a generalized {H}\'enon-{H}eiles {H}amiltonian\fg},
  \emph{J.~Math.~Phys.} (2002), no.~43, p.~1906--1915.

\bibitem{VMC2003}
\bysame , {\og General solution for {H}amiltonians with extended cubic and
  quartic potentials\fg}, \emph{Theor.~Math.~Phys.} (2003), no.~134,
  p.~128--138.

\bibitem{VMC2004a}
\bysame , {\og On reductions of some {KdV}-type systems and their link to the
  quartic {H}\'enon-{H}eiles {H}amiltonian\fg}, in \emph{Bilinear integrable
  systems - from classical to quantum, continuous to discrete, ed.~{P}.~van
  {M}oerbeke}, Kluwer, Dordrecht, 2004.

\bibitem{Woj1984}
{\scshape S.~Wojciechowski} -- {\og Separability of an integrable case of the
  {H}\'enon-{H}eiles system\fg}, \emph{Phys.~Lett.~A} (1984), no.~100,
  p.~277--278.

\end{thebibliography}
\end{document}